\documentclass[aps,prl,twocolumn,groupedaddress,showpacs]{revtex4}
\usepackage{graphicx}
\graphicspath{{figure file fold path}}
\usepackage[dvipsnames,usenames]{color}
\usepackage{color}
\usepackage{amsmath}
\usepackage{float}
\usepackage{amssymb}
\usepackage{amsthm}
\usepackage{mathrsfs}

\emergencystretch=\maxdimen
\begin{document}
\title{Enabling low threshold laser through an asymmetric tetramer metasurface harnessing polarization-independent quasi-BICs}
\author{T. Wang$^{1}$, W. Z. Di$^{2}$, W. E. I. Sha$^{3}$, and R. P. Zaccaria$^{4}$ }

\affiliation{$^1$State Key Laboratory of Integrated Service Networks, Xidian University, Xi'an 710071, China}
\affiliation{$^2$Engineering Research Center of Smart Microsensors and Microsystems of MOE, Hangzhou Dianzi University, Hangzhou, 310018, China}
\affiliation{$^5$College of Information Science and Electronic Engineering, Zhejiang University, Hangzhou 310058, China}
\affiliation{$^4$Istituto Italiano di Tecnologia, via Morego 30, Genoa 16163, Italy}

\date{\today}

\begin{abstract}

We propose and numerically demonstrate a novel strategy to achieve dual-band symmetry-protected bound states in the continuum (BICs) based on a nanodisk tetramer metasurface for lasing generation. The method involves breaking the in-plane symmetry along the diagonal of the metasurface unit cell by introducing air holes in the tetramers. Through our simulations, we show that this flexible approach enables the support of dual-band BICs in the telecom-band range,  with these modes evolving into quasi-BICs with remarkably high quality factors by breaking the symmetry of the system. Furthermore, the ultracompact device exhibits the unique characteristic of being polarization-independent across all viewing angles. Finally, the optically pumped gain medium provides sufficient optical gain to compensate the quasi-BIC mode losses, enabling two mode lasing with ultra-low pump threshold and very narrow optical linewidth in the telecom-band range. Our adaptable device paves the way for polarization-insensitive metasurfaces with multiple lasing resonances. This innovation holds the potential to transform areas like low-threshold lasing and biosensing by delivering improved performance and broader capabilities.

\end{abstract}


\maketitle 

\section{Introduction}
Bound states in the continuum (BICs) are localized states that exist alongside a continuous spectrum of radiating waves, appearing as resonances not interacting with any of the continuum states~\cite{Wang2024, Hsu2016, Hu2021, Romano2019}. Since their first experimental demonstration in 2013, optical BICs have captured significant interest within various photonic systems~\cite{Hsu2013}. Optical BICs are considered dark states with an infinite radiative lifetime, allowing them to achieve infinitely Q resonances and facilitating enhanced interactions between light and matter~\cite{Zhang2024}. This unique characteristic allows for the exploration of distinctive and robust light-matter interactions through the engineering and utilization of BICs~\cite{Gao2023}. Consequently, researchers can investigate topological photonics and create innovative photonic devices with exceptional properties, such as low-threshold lasing and highly sensitive sensing capabilities~\cite{Sun2024, Schiattarella2024, Mohamed2022, Zhang2021}.

Based on the physical mechanisms underlying their formation, BICs can be broadly classified into two groups~\cite{Koshelev2019, Han2024}. The most straightforward type is the symmetry-protected BIC (SP-BIC), which are formed due to the symmetry or separability restricted out-coupling~\cite{Joseph2021, Lee2012}. In periodic photonic structures such as gratings~\cite{Zhang2021}, photonic crystals~\cite{Kang2022}, and even other kinds of metasurfaces~\cite{Fan2023, Campione2016, Liu2019}, the SP-BICs can be found as discrete modes in the Brillouin zone near the gamma ($\Gamma$) point. In theory, ideal BICs cannot be observed because they have zero linewidth and infinite Q-factor in the continuum. However, in real structure or fabricated systems, the Q factor of the BIC is finite due to the restricted dimension, losses, and imperfection in the fabricated system~\cite{Joseph2021}. 
Therefore, only the quasi-BICs can be experimentally realized. Alternatively, people can either slightly break the excitation field symmetry through oblique incidences, or break the in-plane or out-of-plane structural symmetry under normal incidence~\cite{Xiang2020}. In-plane symmetry breaking is a typical approach which involves altering the shape or the structure of the meta-atom, or by introducing a relative tile between paired meta-atoms, whereas the out-of-plane symmetry breaking relies on disturbing the heights of a dimer or trimer cluster~\cite{Feng2023}.

In this paper, we propose a new method to break the symmetry of a tetramer metasurface by introducing air holes for the elements along the inner diagonals. Using this asymmetric array, we numerically explore the excitation of BIC and the evolution of the resulting Fano resonances. Then, we verify the distinct property of the polarization-independence of our design. Finally, by introducing optical gain medium to compensate for system losses, we successfully achieve dual-BIC mode lasing with ultra-low threshold and very narrow optical spectrum linewidth. In particular, the dominating mode of the laser is switched from one state to another when the pump power is higher than a certain value. Therefore, this novel symmetry-breaking technique opens up new possibilities for manipulating the behavior and properties of tetramer metasurfaces, offering promising applications in various fields such as efficient lasing, harmonic generation, biosensing, optical imaging, and entangled photon generation~\cite{Liang2024}.

\section{Structure design}
\begin{figure}[ht!]
\centering
  \includegraphics[width=8.5cm]{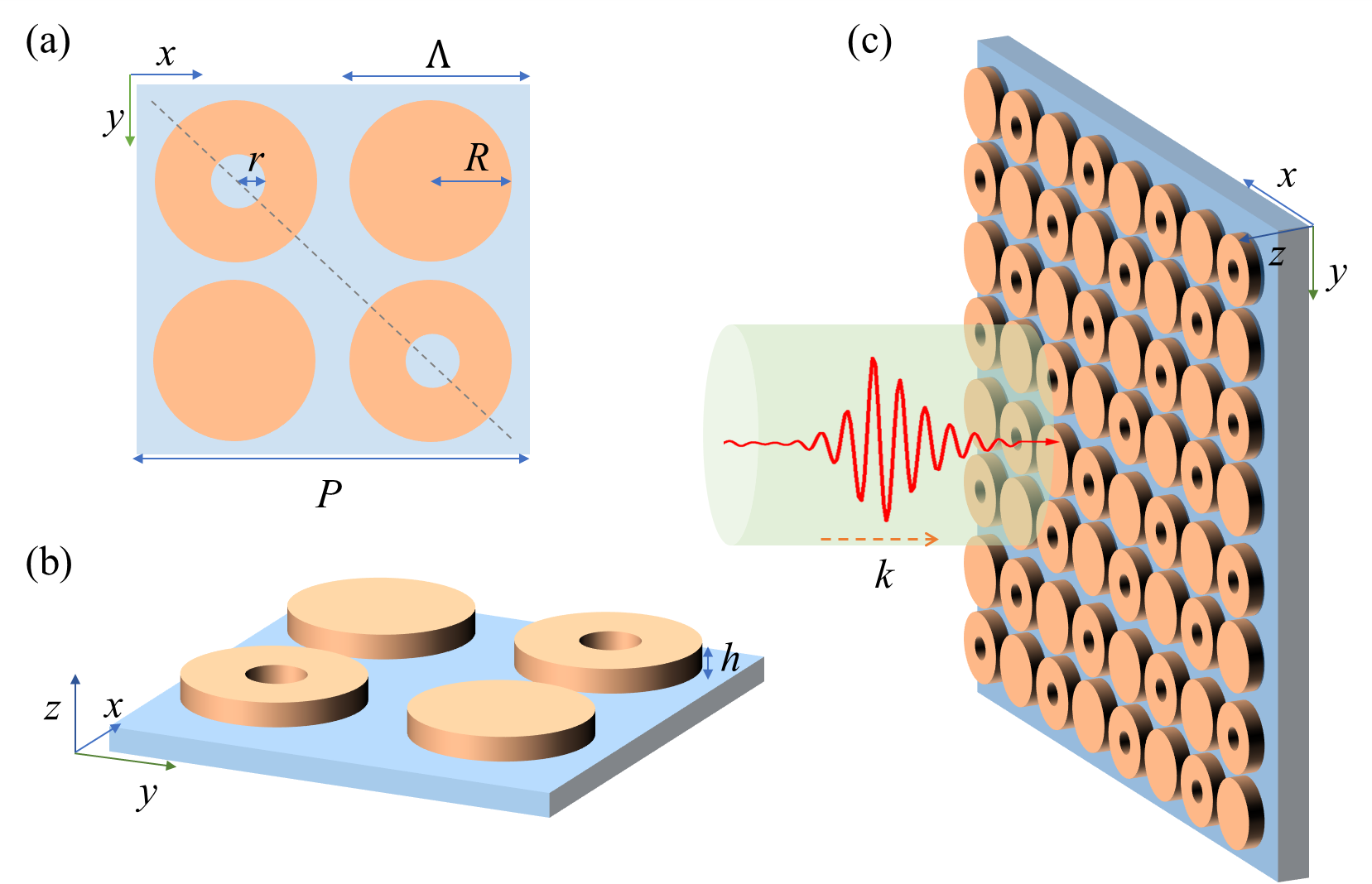}
  \caption{Geometric description of the tetramer cluster: (a) original structure of the unit cell; (b) 3D perspective of a tetramer cluster with in-plane asymmetry; (c) asymmetric tetramer metasurface under optical illumination.}
  \label{Structure}
\end{figure}

Fig.~\ref{Structure} displays the schematic of the proposed structure, which is composed of a square lattice of InGaAsP disk tetramers on a SiO$_2$ substrate. The tetramer cluster is made up of four InGaAsP disks, each having the same geometry parameters and a periodicity of $\Lambda$, as shown in Fig.~\ref{Structure}a. The 3D view of the tetramer cluster is shown in Fig.~\ref{Structure}b, allowing for a more immersive and realistic visualization of the structure. The disks possess radius $R$ and thickness $h$, while the tetramer cluster is defined by periodicity $P=2\Lambda$. To break the in-plane geometric symmetry, an air hole with radius $r$ is introduced in the tetramer cluster as shown in Fig.~\ref{Structure}a. Here, we shall introduce the parameter $\alpha = r / R$ defining the degree of asymmetry of the overall system. 

The quasi-BIC resonance can be detected and studied by analyzing either the transmission or reflection spectrum of the system~\cite{Li2022, Quaranta2018, Yang2023, Novotny2012}. If the system contains quasi-BIC resonances, they will manifest as narrow peaks or dips in the transmission and reflection spectra, indicating resonant modes that enhance or suppress these wave components. This is especially true in the present case, where an all-dielectric structure has been considered, hence providing results free from any broad feature possibly related to the occurrence of absorption. In terms of the adopted computational approach, the finite difference time domain (FDTD) has been selected. In this respect, to guarantee accurate results, the boundary conditions are set to be periodic along the $x$- and $y$-direction to emulate the infinite planar wave and structural periodicity. Vice-versa perfectly matched layer (PML) is employed along the $z$-direction to avoid that the outgoing electromagnetic field could reenter into the structure, hence determining misleading results. As light source was selected a transverse-electric (TE) plane wave impinging on the structure along the $z$-axis (see Fig.~\ref{Structure}c) having the electric field vector oriented along the $y$-direction.

The tetramer cluster chosen for the present analysis is characterized by a period $P$ = 1200 nm and a height $h$ = 250 nm. Furthermore, the refractive index associated with the InGaAsP disks and the SiO$_2$ substrate is set to 3.47 and 1.45, respectively. Importantly, the proposed structure can be experimentally fabricated owing to the well established fabrication techniques involving InGaAsP, a commonly used semiconductor material. In particular, electron-beam lithography (EBL) and molecular-beam epitaxy (MBE)~\cite{Hwang2021, Kodigala2017}, can be employed to realize the structure shown in Fig.~\ref{Structure}. Furthermore, the bandgap of InGaAsP can be tailored to specific wavelengths, thereby making this material one of the best candidates for the development of highly efficient lasers, such as edge-emitting lasers and vertical-cavity surface-emitting lasers (VCSELs)~\cite{Yang1990, Behrend1998}.

\section{Results and discussions}
\subsection{\textit{Evolution of Fano resonance dominated by SP-BIC}}
\begin{figure}[ht!]
\centering
  \includegraphics[width=8.5cm]{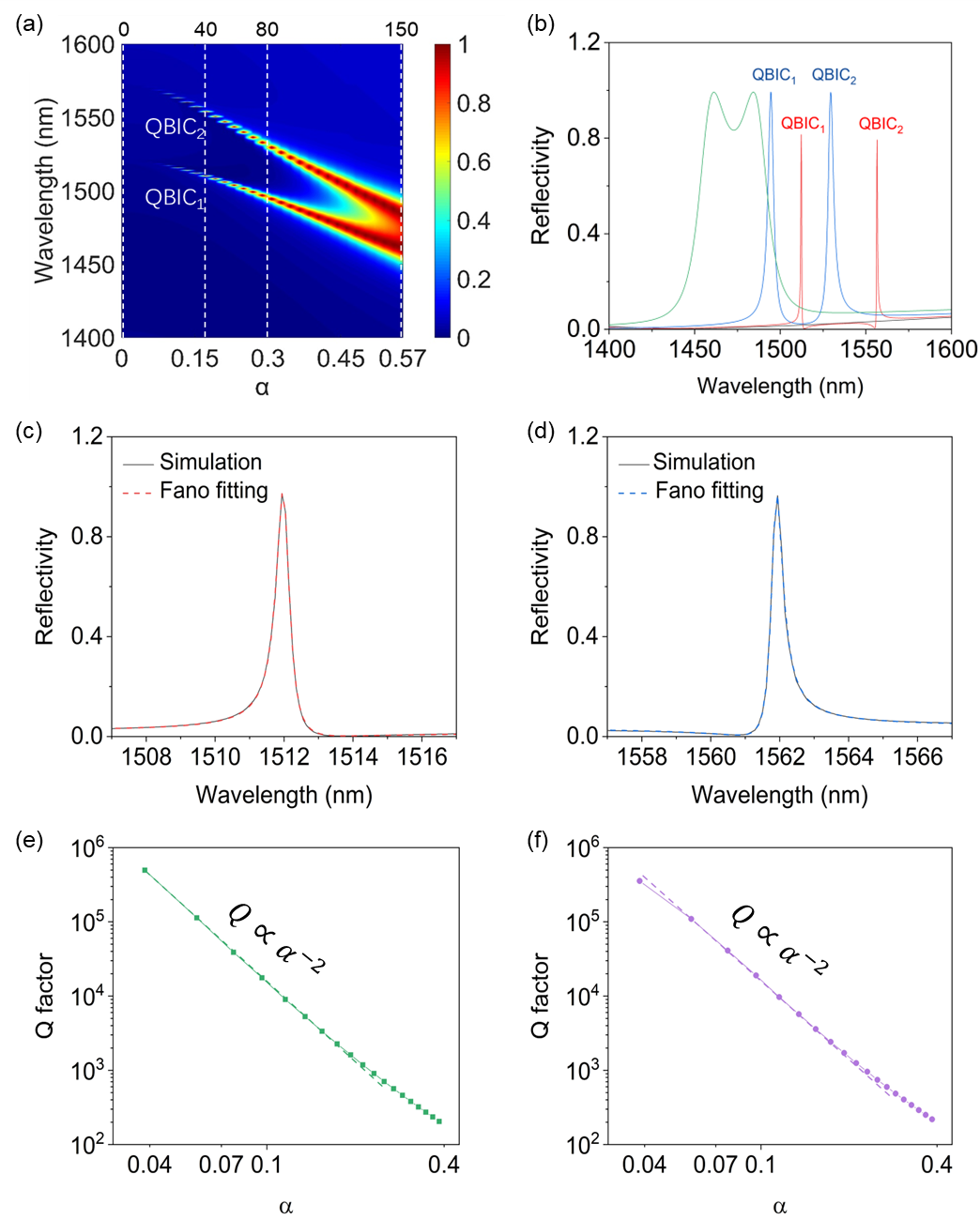}
  \caption{(a) Reflection spectra mapping of the parameter $\alpha$ (bottom)/air holes radius $r$ (top). The four dash lines correspond to the curves in (b); (b) numerically calculated reflection spectrum when $\alpha =$ 0 (black), 0.15 (pink), 0.32 (blue) and 0.58 (green); (c) and (d) Fano fitting of BIC$_1$ and BIC$_2$ for $\alpha = $0.15; (e) and (f) describe the quality factor versus the asymmetry parameter $\alpha$ for (e) BIC$_1$ and (f) BIC$_2$ in log-linear scale.}
  \label{spectra-mapping-1}
\end{figure}

Fig.~\ref{spectra-mapping-1}a shows the mapping of the reflection spectra while changing the radius $r$ from 0 to 150 nm (correspondingly, $0 \leq \alpha \leq 0.58$) while keeping R constant at 260 nm. When $\alpha = 0$, the reflection spectrum in the range of 1400 to 1600 nm is close to zero, therefore presenting a high transmittance (none of the adopted materials is lossy). This result can be attributed to the presence of SP-BICs, which display resonances with infinitely narrow linewidth~\cite{Zhang2018, Doeleman2018}. This is further confirmed by the observation that, as $\alpha$ increases, the symmetry of the structure is disturbed and two resonances gradually appear in the reflection spectrum. This particular phenomenon can be attributed to the coupling of bound states to radiative channels, resulting in the excitation of two quasi-BICs. These modes have been marked by QBIC$_1$ and QBIC$_2$, the latter with a longer emission wavelength. 

Interestingly, we notice a blue-shift of both BIC$_1$ and BIC$_2$ upon increase of $\alpha$, possibly due to the decrease of the effective refractive index of the structure~\cite{Zhou2018, Lustig2023}. At the same time, their line widths become larger, indicating the degradation of the Q-factors of these two quasi-BIC modes. Finally, when the size of the air hole is increased even further, these two modes merge into one another, resulting in a phenomenon known as ``degeneracy''. This implies that the characteristics of the two individual modes become indistinguishable or interchangeable. This merging of the modes can occur due to the increasing similarity in their respective wavelengths, allowing them to overlap and share common features~\cite{Wang2020}. Fig.~\ref{spectra-mapping-1}b shows the representative reflection spectra of the metasurface of the tetramer when $\alpha = $0 (black line), 0.15 (pink line), 0.32 (blue line) and 0.58 (green line), which confirms the analysis of Fig.~\ref{spectra-mapping-1}a.

To confirm the resonance and calculate the Q-factor of BIC modes $Q = \omega_0/2\gamma$, we employ the conventional Fano formula~\cite{Saadatmand2023}:

\begin{eqnarray}
 T_{fano} (\omega) =  \vert a_1 + ja_2 + \frac{b}{\omega - \omega_0 + j \gamma}\vert^2 \, ,  
\end{eqnarray}

\noindent where $\omega_0$ represents the resonant frequency, a$_1$, a$_2$, and b are constants, and $\gamma$ indicates the total damping rate and characterizes the Q-factor of the QBIC modes. The fitting reflection spectra for QBIC$_1$ and QBIC$_2$ modes, corresponding to $\alpha$ = 0.15, are shown
in Fig.~\ref{spectra-mapping-1}c and d, respectively. Specifically, QBIC$_1$ exhibits a resonance at 1511 nm with a Q-factor of 6946, while QBIC$_2$ is centered at 1562 nm and possesses a Q-factor of 7800. These results are consistent with the numerical simulation outcomes, which are illustrated by the black solid lines in Fig.~\ref{spectra-mapping-1}c and d, respectively.

Fig.~\ref{spectra-mapping-1}e and f show the dependence of the Q factor of the two quasi-BIC modes on the asymmetry parameter $\alpha$ (in log-log scale). The Q factors of these two resonant modes become very similar when $\alpha > 0.4$, suggesting a mode degradation possibly leading to a loss of confinement. This is confirmed by the green line in Fig. \ref{spectra-mapping-1}b ($\alpha= 0.58$), clearly showing two very broad linewidths associated to the two resonance modes. For this reason, here we shall focus on the region where $\alpha < 0.4$. Furthermore, the two figures further stress the enhanced radiation loss occurring for high alpha values, by showing a steady Q factor decrease upon $\alpha$ increase. When $\alpha = 0$, both modes are BIC resonances and exhibit infinite Q factors. When a degree of asymmetry $\alpha = 0.1$ is introduced, the Q factors of both QBIC$_1$ and QBIC$_2$ remain high with values above $10^4$. However, for a larger degree of asymmetry $\alpha = 0.2$, the Q factor substantially decreases to $10^3$, suggesting an augmented radiation loss of the tetramer metasurface. Furthermore, our study reveals that the Q factors of both modes exhibit a clear dependence on $\alpha$ following an inverse quadratic law when $\alpha > 0.2$, which is in line with the findings of previous studies~\cite{Koshelev2018, Liu2019}, hence suggesting to be in the presence of SP-BICs.

\begin{figure}[ht!]
\centering
  \includegraphics[width=8.5cm]{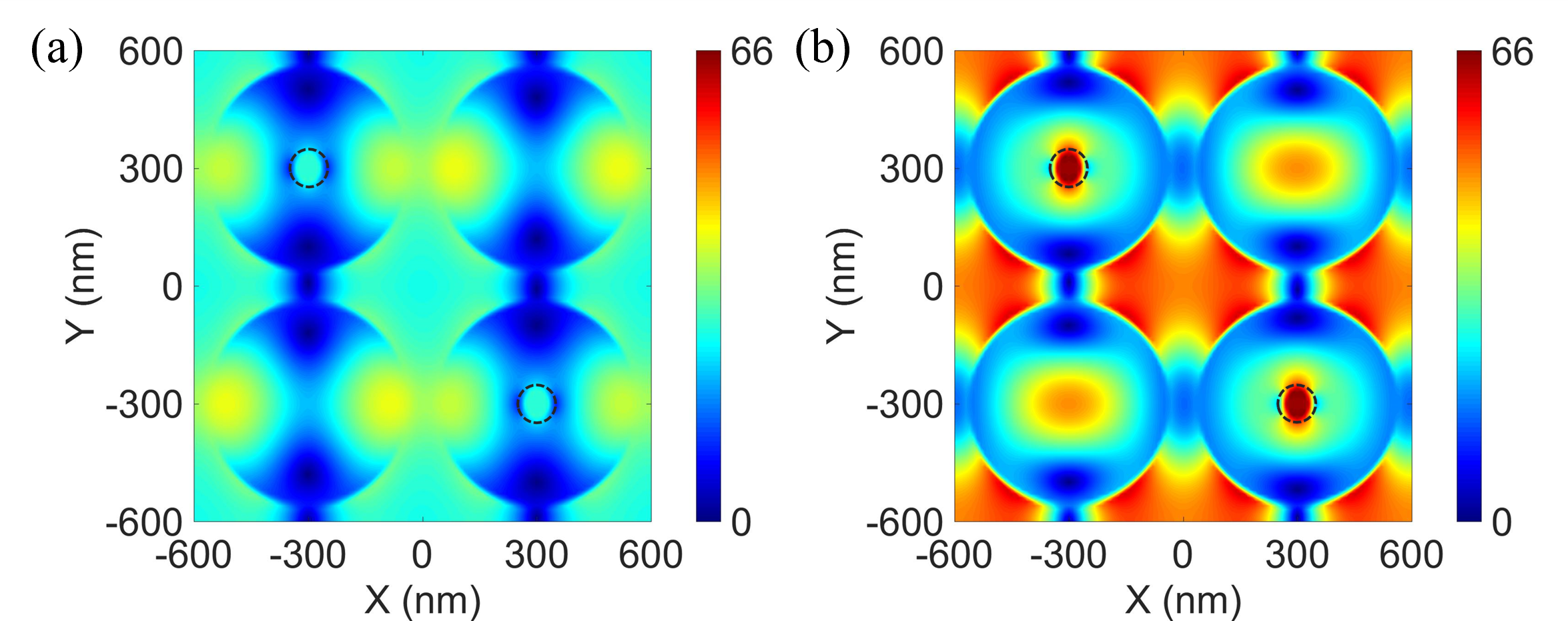}
  \caption{Electromagnetic field distribution profiles of two quasi-BIC modes when $\alpha = 0.15$: (a) QBIC$_1$; (b) QBIC$_2$. These plots were calculated at $z=$ 125 nm, corresponding to the half height of the InGaAsP disks.}
  \label{field}
\end{figure}

Fig.~\ref{field} shows the simulated intensity distributions of the electromagnetic field of QBIC$_1$ and QBIC$_2$ in the ($x$, $y$)-plane, providing a clear visual representation of the spatial distribution of both modes under asymmetric conditions ($\alpha = 0.15$). This visualization provides a better understanding of the resonances QBIC$_1$ and QBIC$_2$. As shown in Fig.~\ref{field}a, for the QBIC$_1$ mode we observe high electromagnetic intensity especially within the InGaAsP disks, with a distribution clearly resulting from a x-polarized impinging radiation. Conversely, for the QBIC$_2$ mode, the electromagnetic radiation is primarily concentrated within the air holes (highlighted by dashed lines) and in the air region surrounding the InGaAsP disks. As already depicted in Fig. \ref{spectra-mapping-1}a, this results in the QBIC$_1$ resonance being blue-shifted compared to the QBIC$_2$ resonance (InGaAsP refractive index $> 1$)~\cite{Vaity2021, Algorri2024}.

To better understand the properties of QBIC modes, the cartesian multipole decomposition method is employed to compute the scattering power for diverse multipole excitations. By integrating the displacement current density $j(r)$ across the unit cell, a comprehensive portrayal of the far-field responses of electromagnetic sources is achieved. Subsequently, for a specified frequency, the total far-field scattered power is determined for all multipoles. Among these, the eight multipoles including the electric dipole (ED), the magnetic dipole (MD), the magnetic toroidal dipole (MTD), the electric toroidal dipole (ETD), the electric quadrupole (EQ), the magnetic quadrupole (MQ), the electric toroidal quadrupole (ETQ), and the magnetic toroidal quadrupole (MTQ), are defined by the following equations respectively~\cite{Saadatmand2023, Tang2017}:

\begin{eqnarray}
 &P = \frac{1}{\omega}\int jd^3r \, ,  \\
 &M = \frac{1}{c}\int (r \times j)d^3r \, ,  \\
 &T = \frac{1}{10c} \int [(r \cdot j)r - 2r^2j]d^3r \, ,  \\
 &T^e = \frac{\omega^2}{20c^2} \int [(r \times j)_{\alpha}  r^2]d^3r \, ,  \\
 &Q^e_{\alpha \beta} = \frac{1}{2i\omega} \int [r_{\alpha}j_{\beta} + r_{\beta}j_{\alpha} - \frac{2}{3} (r \cdot j) \delta_{\alpha, \beta}]d^3r \, ,  \\
 &Q^m_{\alpha \beta} = \frac{1}{3c} \int [(r \times j)_{\alpha} r_{\beta} + (r \times j)_{\beta}r_{\alpha}]d^3r \, ,  \\
 &Q^{ET}_{\alpha \beta} = \frac{i\omega}{42} \int r^2 [r_{\alpha} (r \times j)_{\beta} + (r \times j)_{\beta}r_{\alpha}]d^3r \, ,  \\
 &Q^{MT}_{\alpha \beta} = \frac{i\omega}{28c} \int [4r_{\alpha}r_{\beta} (r \cdot j)  - 5r^2(r_{\alpha}j_{\beta} + r_{\beta}j_{\alpha}) \\ & +  \nonumber 2r^2\delta_{\alpha\beta}(r \cdot j)]d^3r \, ,
\end{eqnarray}

\noindent where $\omega$ represents the angular frequency, $r$, and $c$ denote the position vectors and the speed of light, respectively. $\delta$ is the Dirac delta function, and the subscripts $\alpha$, $\beta$ take values $x$, $y$, $z$. The scattered power associated with each multipole moment can be calculated from Ref.~\cite{Huang2012}. They are expressed as

\begin{eqnarray}
 I_P = \frac{2\omega^4}{3c^3} \vert P \vert^2 \, ,  \\
 I_M = \frac{2\omega^4}{3c^3} \vert M \vert^2 \, ,  \\
 I_T = \frac{2\omega^6}{3c^5} \vert T \vert^2 \, ,  \\
 I_{EQ}= \frac{\omega^6}{5c^5} \vert Q^e_{\alpha \beta} \vert^2 \, , \\
 I_{ET} = \frac{2\omega^6}{3c^5} \vert Q^{ET}_{\alpha \beta} \vert^{2} \, ,  \\
 I_{MQ} = \frac{\omega^6}{20c^5} \vert Q^m_{\alpha \beta} \vert^2 \, ,
\end{eqnarray}

\noindent where $I_P$, $I_M$, $I_T$, $I_{EQ}$, $I_{ET}$, and $I_{MQ}$ denote the scattering power from the electric dipole, the magnetic dipole, the toroidal dipole, the electric quadrupole, the electric toroidal quadrupole, and the magnetic quadrupole, respectively~\cite{Huang2012}.

\begin{figure}[ht!]
\centering
  \includegraphics[width=8.0cm]{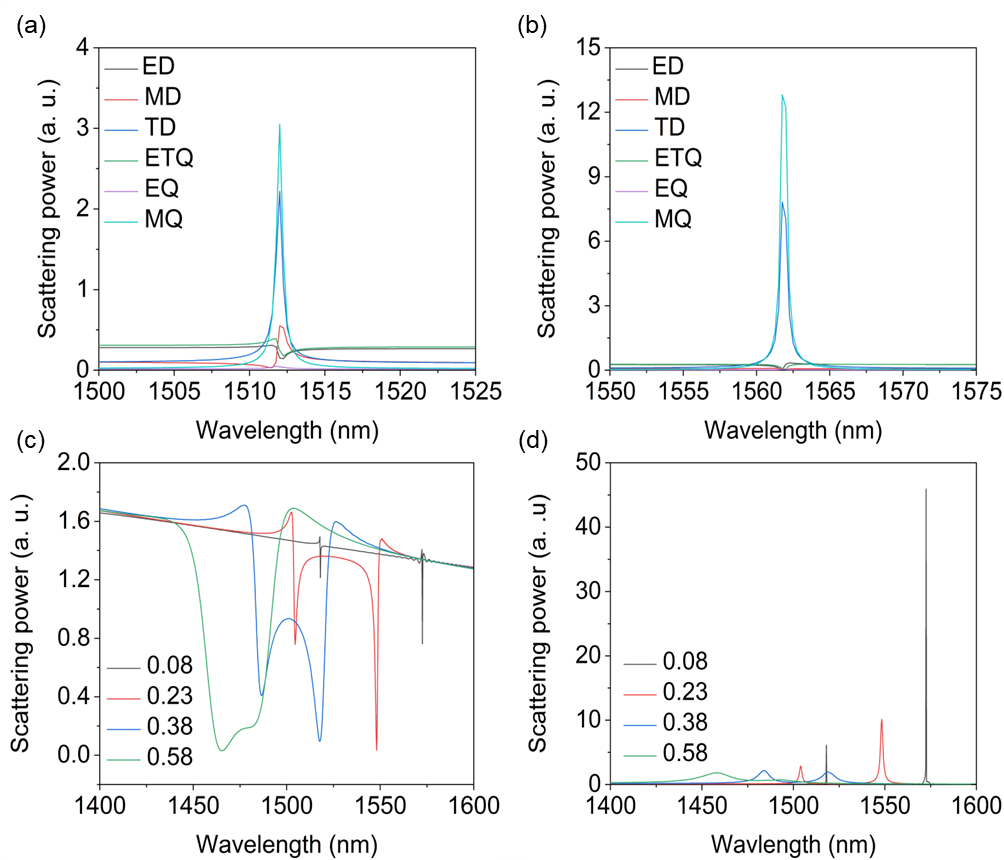}
  \caption{Far field scattering power of the QBIC$_1$ mode from electric dipole ($I_P$, denoted by ED), magnetic dipole ($I_M$, denoted by MD), toroidal dipole ($I_T$, denoted by TD), electric toroidal quadrupole ($I_{ET}$, denoted by ETD), electric quadrupole dipole ($I_{EQ}$, denoted by EQ), magnetic quadrupole dipole ($I_{MQ}$, denoted by MQ);  (b) The same as (a) but for the QBIC$_2$ mode, respectively; (c) and (d) the contributions of ED and MQ to both modes, the mode at short wavelength is the QBIC$_1$ mode, and the one at longer wavelength is the QBIC$_2$ mode}
  \label{dipoleanlaisi}
\end{figure}

The multipole decompositions of the structure for alpha=0.15 are depicted in Fig.~\ref{dipoleanlaisi}. Specifically, Fig.~\ref{dipoleanlaisi}a illustrates that for QBIC$_1$ mode, the MQ is the dominant multipole, accompanied by partial contributions from the TD and MD. Notably, the contributions from other higher-order multipoles are significantly suppressed in this instance. Analogously, QBIC$_2$ mode presents a similar property, with the MTQ remaining the dominant multipole and the TD contributing partially. In this case as well, the contributions from other higher-order multipoles are negligible, as shown in Fig.~\ref{dipoleanlaisi}b.

The multipole analysis provides insights into the underlying mechanisms responsible for the dark mode behaviour associated with QBIC$_1$ and QBIC$_2$ at $\alpha$ = 0 and for small $\alpha$ values (see Fig.~\ref{spectra-mapping-1}a). Specifically, from Fig. Fig.~\ref{dipoleanlaisi}c, the evolution of the electric dipole (ED) multipole to QBIC$_1$ and QBIC$_2$ with changing $\alpha$ is evident. The analysis reveals that the ED multipole increases with 
$\alpha$, which explains why a linearly polarized incident radiation cannot effectively excite the structure at $\alpha$ = 0 or for small $\alpha$ values, as observed in Fig.~\ref{spectra-mapping-1}a. This is because linearly polarized radiation corresponds to an ED-like mode. Considering that, in addition to energy and momentum parallel to the structure surface, the symmetry of the incident radiation must also be conserved during mode transformation (such as when an incident radiation excites a mode), it becomes apparent that a linearly polarized radiation can best excite ED modes~\cite{Algorri2022}. In turn, this means that a linearly polarized radiation can properly couple to the structure shown in Fig.~\ref{Structure} only at high $\alpha$ values. This finding is confirmed by Fig.~\ref{dipoleanlaisi}d, where the evolution of the MQ multipole is shown. The results say that at $\alpha$ = 0 the main multipole is MQ (see Fig.~\ref{dipoleanlaisi}a and Fig.~\ref{dipoleanlaisi}b), a contribution that decreases by increasing $\alpha$, exactly the opposite found for the ED multipole.

\begin{figure}[ht!]
\centering
  \includegraphics[width=8.5cm]{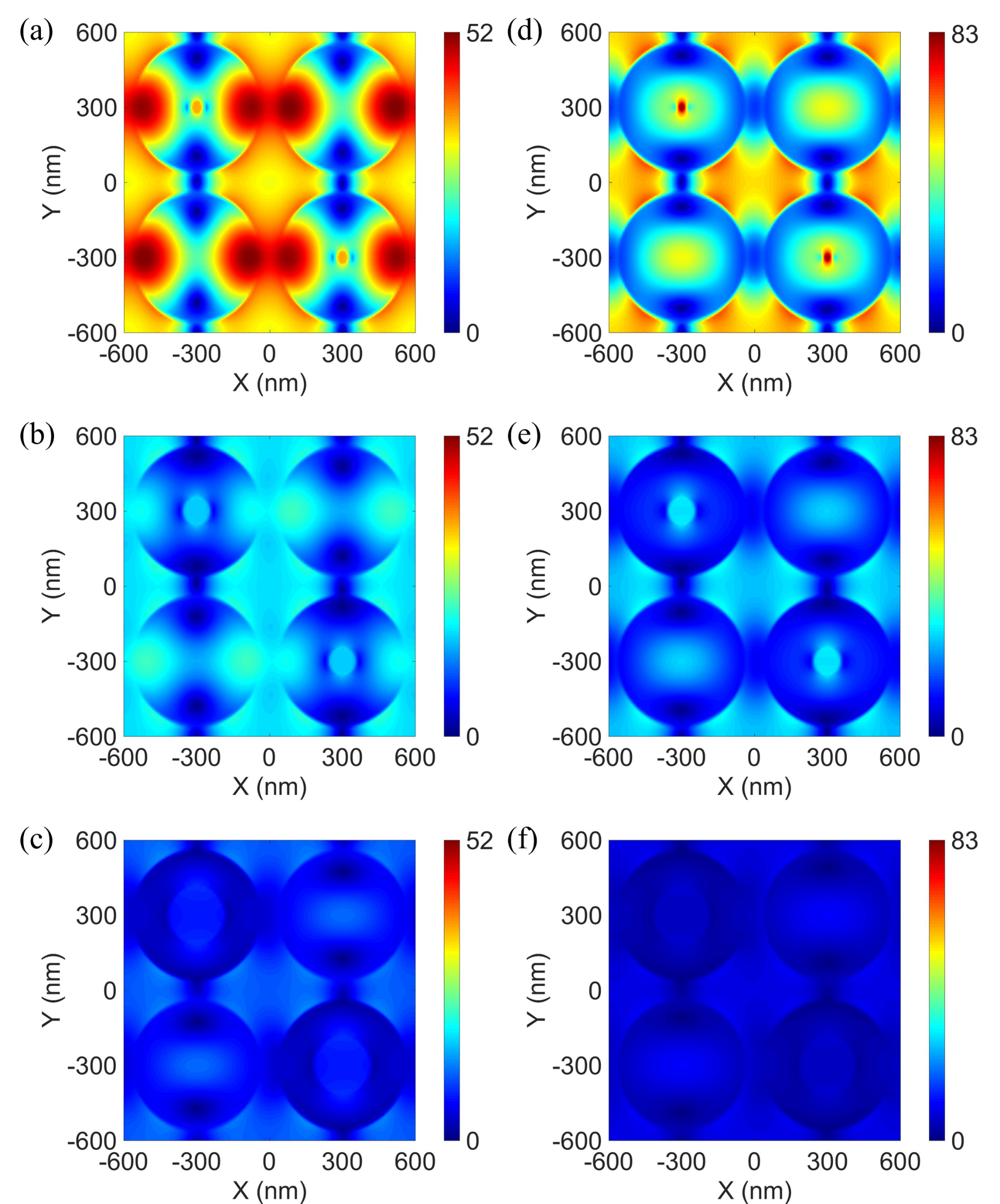}
  \caption{Electromagnetic field distribution profiles of QBIC$_1$ and QBIC$_2$ modes when $\alpha = 0.08$ (a) and (d), 0.23 (b) and (e), 0.39 (c) and (f) , respectively; (a)-(c) for QBIC$_1$ mode; (d)-(f) for QBIC$_2$ mode.}
  \label{compare}
\end{figure}

In order to gain an even deeper understanding of the impact of the asymmetry, we further conduct an analysis on the electromagnetic field distributions of the QBIC$_1$ and QBIC$_2$ modes in ($x$, $y$)-plane for the structures with air hole size of $\alpha =$ 0.08, 0.23, and 0.39. As shown in Fig.~\ref{compare}a-c, we find that the maximum intensity of QBIC$_1$ mode decreases with the size of the air holes (similar behaviour occurs also for QBIC$_2$, as shown in Fig.~\ref{compare}d-f). The decreasing trend can be attributed to the reduction in the factor $Q$ of the QBIC$_1$ mode as already observed in Fig.~\ref{spectra-mapping-1}c and d. Indeed, a lower $Q$-factor implies a weaker confinement for the electromagnetic field within the structure, thus resulting in the observed weakening of the field intensity. Furthermore, when comparing Fig.~\ref{compare}c and f, it is observed that the two resonances tend to converge towards a similar pattern, namely QBIC$_1$ and QBIC$_2$ somehow merge in a unique kind of mode, a behaviour matching with the results of Fig.~\ref{spectra-mapping-1}a.

\subsection{\textit{Polarization independent behaviour}}
\begin{figure}[ht!]
\centering
  \includegraphics[width=8.5cm]{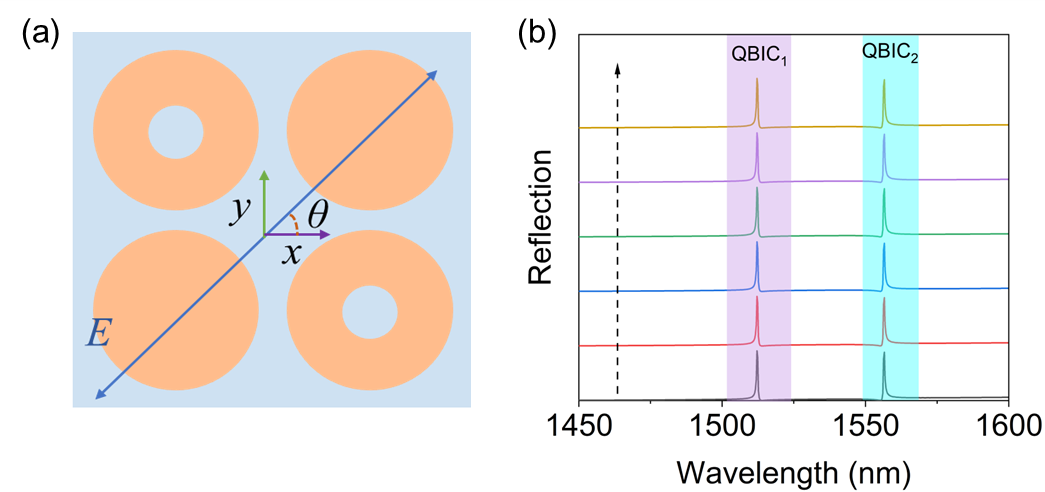}
  \caption{(a) Definition of the polarization angle $\theta$ of the excitation wave; (b) Reflection spectra for different polarization angles, where the polarization angle increases as indicated by the dashed arrow. Starting from bottom and moving towards the top, the curves on the graph represent the reflection spectra for polarization angles of 0$^\circ$, 30$^\circ$, 45$^\circ$, 60$^\circ$, 75$^\circ$, and 90$^\circ$.}
  \label{Resp}
\end{figure}

To verify the polarization independent behaviour of the quasi-BIC resonances in the metasurface, it was calculated their reflection spectra with different excitation angles and for $\alpha =$ 0.15. The direction of polarization of the incident linearly polarized beam is defined by the angle ($\theta$) between the incident electric field and the $x$-axis (see Fig.~\ref{Resp}a). As shown in Fig.~\ref{Resp}b, when the incident beam is polarized along the $x$ axis (i.e., $\theta = 0^\circ$), the resonant wavelengths of QBIC$_1$ and QBIC$_2$ are 1512 nm and 1557 nm, respectively. As the polarization angle $\theta$ varies from $0^\circ$ to $90^\circ$ (see dashed arrow in the figure), the resonant wavelengths of both quasi-BIC modes remain nearly unchanged. This observation clearly indicates that the proposed metasurface shows polarization insensitivity when exciting the quasi-BIC resonances. This unique characteristic can be exploited to reduce, if not eliminate, the need for polarization calibration, thus offering greater flexibility for practical applications such as wireless communication systems, radar technology, and optical sensing~\cite{Cui2024, Zhang2023}. 

\subsection{\textit{Dual mode lasing and characterization}}
Lasing action based on BICs or quasi-BICs in two-dimensional metasurface structures has recently garnered significant attention in the field of photonics~\cite{Kodigala2017, Zhang2022, Heilmann2022}. These unique structures exhibit remarkable properties that enable the confinement and manipulation of light at the nanoscale, leading to potential advancements in lasers and optical devices~\cite{Pavlov2018}. In the present context, the selected structure for realizing lasing from both the quasi-BIC modes is defined by $\alpha$ = 0.15. Furthermore, since InGaAsP is a natural gain material, the imaginary part of its refractive index is considered through all the numerical simulations.


The gain medium can be well described as a four-energy-level system, and a semi-quantum framework is adopted to simulate the interaction between the electromagnetic fields and gain medium. Details about the population dynamics of each energy level within the gain medium, including the ground state and the three excited states, can be found in the Ref.~\cite{Di2023, Zhang2022}. Similarly to the calculation of the quasi-BIC modes, the FDTD method is employed also for simulating the lasing effect, where a 4 ps plane wave pump pulse is deployed on the metasurface along $z$-direction. The detailed parameters used in the simulation are described in~\cite{Di2023}.

\begin{figure}[ht!]
\centering
  \includegraphics[width=5.0cm]{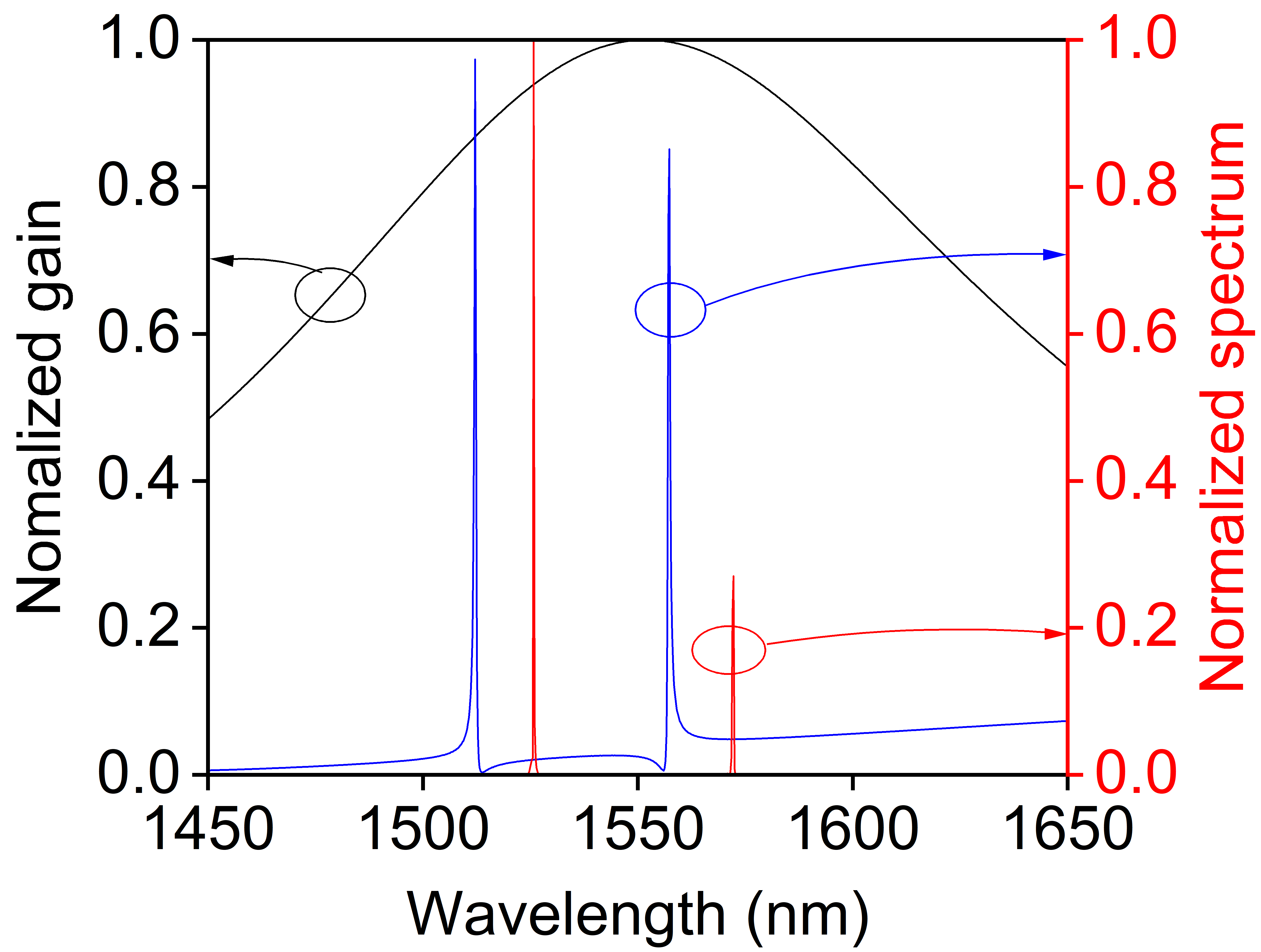}
  \caption{Gain spectrum of InGaAsP (black curve), QBICs reflection peaks (blue curve), and lasing emission of the QBIC modes (red curve) operated above threshold.}
  \label{gainlasing}
\end{figure}

Fig.~\ref{gainlasing} shows the gain spectrum of the designed metasurface (black curve), the QBIC modes reflection spectra (blue curve) and their lasing emission (red curve), the latter being red-shifted against the reflection spectra due to the quantum efficiency of the gain medium. It is evident that the gain spectrum exhibits a wide linewidth, and its central region overlaps the peaks associated to the QBIC modes. So, this characteristic can guarantee that cavity losses are effectively compensated, enabling optimal laser operation~\cite{Wang2017, Sun2021}. 

\begin{figure}[ht!]
\centering
  \includegraphics[width=8.5cm]{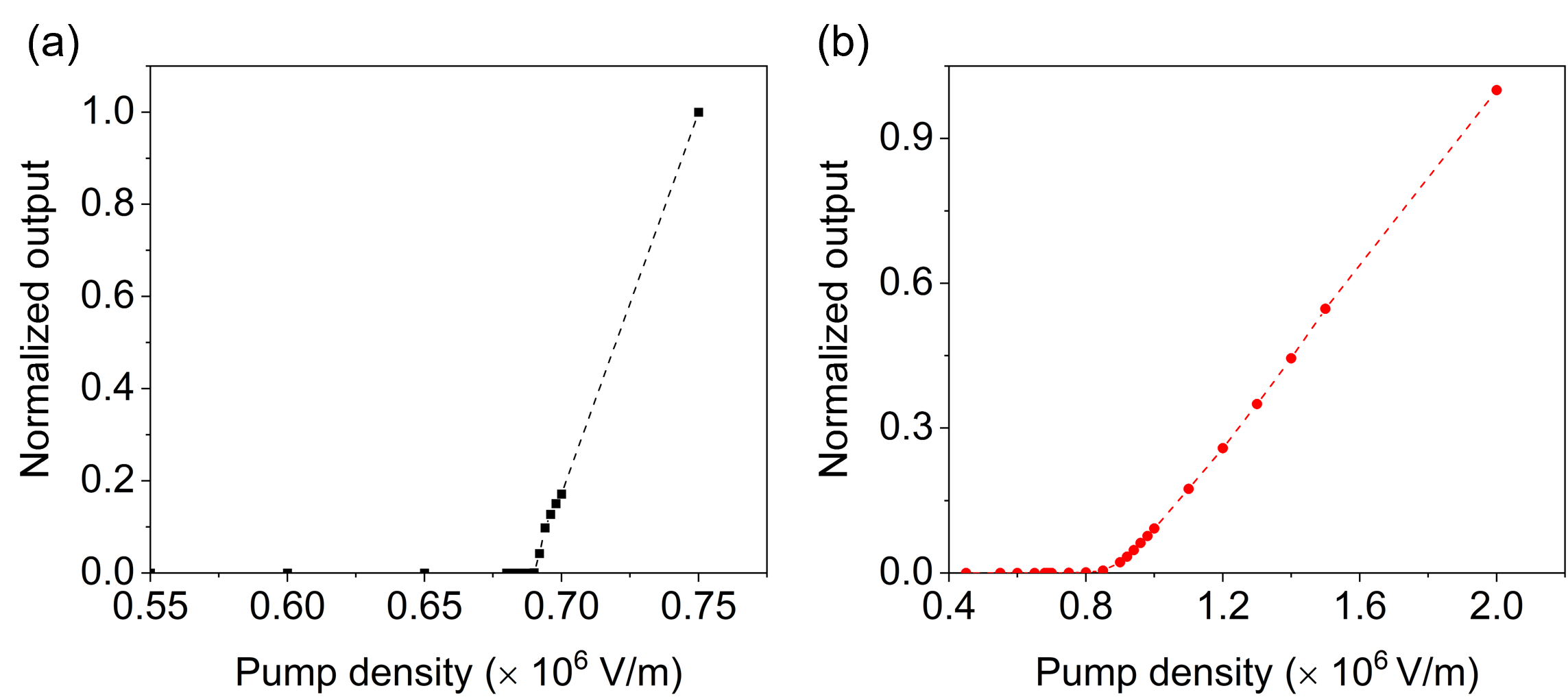}
  \caption{Lasing function curves of two quasi-BIC modes: (a) QBIC$_1$; (b) QBIC$_2$.}
  \label{lasing}
\end{figure}

Fig.~\ref{lasing} shows the normalized (i.e., divided by the maximum output value) input-output function curves of the QBIC$_1$ (a) and QBIC$_2$ (b) modes, providing a comprehensive understanding of their characteristic lasing behaviors. From these results, it is clear that the QBIC$_1$ mode has a significantly lower lasing threshold compared to the QBIC$_2$ mode, indicating that the QBIC$_1$ mode requires substantially less input power to initiate the lasing action. In addition, both modes exhibit a consistent and proportionate growth with input power within the specified power range. This implies that the output of the system experiences a steady and uniform increase as the input power surpasses a certain threshold. The linear increment in output power enhances the overall performance of both modes and reaffirms their reliability and effectiveness in practical applications.

\begin{figure}[ht!]
\centering
  \includegraphics[width=5.5cm]{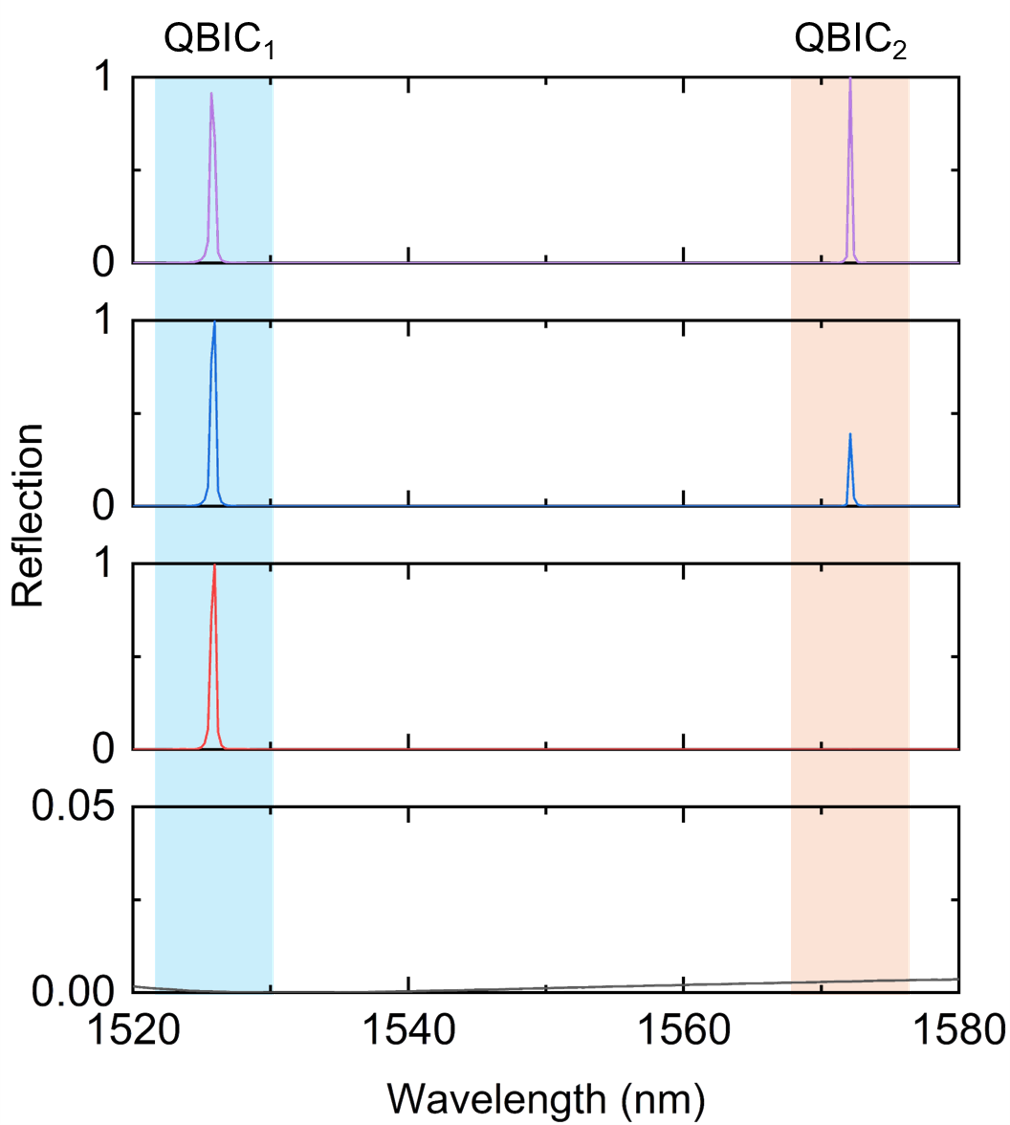}
  \caption{Representative spectra of the device as function of input power: from bottom to the top, the pump are $4.5 \times 10^5$, $8.0 \times 10^5$, $1.5 \times 10^6$, and $2.5 \times 10^6$ V/m, respectively.}
  \label{lspectra}
\end{figure}

Fig.~\ref{lspectra} presents the variation of lasing emission peaks with respect to pump power. These spectra provide valuable insights into the dynamic evolution of the lasing behavior of QBIC$_1$ and QBIC$_2$ modes. Specifically, when the laser operates below threshold at a pump power of $4.5 \times 10^5$ V/m, no discernible mode is observed. As the pump power increases to $8.0 \times 10^5$ V/m, the peak describing the lasing effect associated to the QBIC$_1$ mode can be clearly observed, whereas no lasing effect associated to the QBIC$_2$ mode is noticed due to the insufficient input power. The result is consistent with input-output function curves in Fig.~\ref{lasing}. As the pump power is further elevated to $1.5 \times 10^6$ V/m, the QBIC$_2$ mode starts lasing as well, yet the QBIC$_1$ mode remains the dominant mode, with its intensity continuously increasing and linewidth steadily decreasing (as indicated by the blue curve, representing the spectrum of the device at a pump power of $1.5 \times 10^6$ V/m). Finally, when the pump power reaches $2.5 \times 10^6$ V/m, the intensity of the QBIC$_2$ mode surpasses that of the QBIC$_1$ mode, signifying a mode-switching event.

%
%
%

\section{Conclusions}
In conclusion, we have demonstrated a polarization-independent laser emitter, working in the telecom-band, based on a tetramer metasurface capable of sustaining asymmetric BICs. This designed metasurface consists of four dielectric nanodisks with air holes positioned along the diagonal of the tetramer cluster, showing a symmetry-breaking effect in both vertical and horizontal directions. By gradually increasing the air hole diameter, two high Q-factor quasi-BIC modes were successfully generated. Finally, the effect of a gain medium inserted in the metasurface was analyzed in terms of lasing behaviour. Indeed, it was shown that by compensating the tetramer metasurface losses, an ultracompact laser device emitting in the telecom-band was achieved. In particular, the laser could operate at two different frequencies, corresponding to the two BICs modes sustained by the tetramer metasurface. This result is a significant achievement in the field of laser technology, as it opens up new possibilities for compact and efficient laser devices in the telecom-band range.

\section*{Acknowledgment}
The authors would like to thank Prof. Yang Li and Dr. Yueyang Liu for their insightful discussion. T. W. was supported by Key Research and Development Plan of Shaanxi Province, China (Grant No. 2024GH-ZDXM-42), National Natural Science Foundation
of China (Grant No. 61804036). R. P. Z. was supported by  National Natural Science Foundation of China (Grant No. 32071317). 



\end{document}